\documentclass[%
 reprint,
nofootinbib,
 amsmath,amssymb,
 aps,
]{revtex4-2}

\usepackage{graphicx}
\usepackage{bm}
\usepackage{amsthm}
\usepackage{amsmath}
\usepackage{hyperref}
\usepackage{xspace}
\usepackage{cleveref}
\usepackage{natbib}

\hypersetup{
    pdfauthor={Andrew Fowlie},
    colorlinks=true,
    allcolors=[rgb]{0.0, 0.2, 0.7},
}

\usepackage[english]{babel}
\usepackage[T1]{fontenc}
\usepackage[utf8]{inputenc}
\usepackage{lmodern}

\newcommand{\like}{\mathcal{L}}
\newcommand{\pvalue}{\textit{p-}value\xspace}
\newcommand{\halfchisq}{\ensuremath{\mbox{\footnotesize\ensuremath{\frac12}}\chi^2}}
\newcommand{\gev}{\,\text{GeV}}
\newcommand{\ccite}{ref.~\cite}
\newcommand{\Ccite}{Ref.~\cite}

\defcitealias{Crivellin:2021ubm}{``Accumulating Evidence for the Associate Production \\ of a Neutral Scalar with Mass around 151 GeV''}

\begin{document}

\title{Comment on \citetalias{Crivellin:2021ubm}}

\author{Andrew Fowlie}
\email{andrew.j.fowlie@NJNU.edu.cn}
\affiliation{%
Department of Physics and Institute of Theoretical Physics, Nanjing Normal University,
Nanjing, Jiangsu 210023, China
}%

\begin{abstract}
A recent paper \cite{Crivellin:2021ubm} accumulates evidence for a new fundamental particle by combining several CMS and ATLAS searches for the Standard Model Higgs boson. The putative particle is a neutral scalar, $S$, with a mass of about $151\gev$. The reported significances are $5.1\sigma$ local and $4.8\sigma$ global. This nearly reaches the $5\sigma$ threshold for a discovery in high-energy physics. In this brief note we cast doubt on the strength of the evidence for a new particle. After taking into account the fact that signals were fitted to six different channels, we find that the significances are only $4.1\sigma$ local and $3.5\sigma$ global. The code and instructions for reproducing our calculations are available at \href{https://github.com/andrewfowlie/accumulating_evidence}{this github repo}.
\end{abstract}

\maketitle

\section{Introduction}

In \ccite{Crivellin:2021ubm} several ATLAS and CMS resonance searches are combined into six channels. In each channel, the background model predicts a smooth non-resonant spectrum parameterised by four nuisance parameters, collectively denoted by $\boldsymbol\theta$. The signal model, on the other hand, predicts a Crystal-ball shaped resonance in each channel parameterised by seven parameters: a single mass parameter, $m$, common to each channel, and six positive signal strength parameters $\mu_i \ge 0$ for $i = 1 \text{ -- } 6$, one for each channel. 

\Ccite{Crivellin:2021ubm} construct a test-statistic based on a profiled likelihood ratio for the background model and background plus signal model,
\begin{equation}\label{eq:lambda}
    \lambda \equiv -2 \log{\frac{
        \max_{m,\boldsymbol{\mu}, \boldsymbol\theta} \like(m, \boldsymbol{\mu}, \boldsymbol\theta)
        }{
        \max_{\boldsymbol\theta} \like(\boldsymbol{\mu} = 0, \boldsymbol\theta)
        }},
\end{equation}
where $\like(m, \boldsymbol{\mu}, \boldsymbol\theta)$ is the likelihood of the observed data for the model with parameters $m$, $\boldsymbol{\mu}$ and $\boldsymbol\theta$, $\boldsymbol{\mu} = 0$ corresponds to the background model, and the maximisation over the signal strengths is constrained by $\mu_i \ge 0$. \Ccite{Crivellin:2021ubm} find $\lambda = 26.01$~\cite{email}; we haven't verified this result, but assume that it is correct. This test-statistic is used to compute local and global significances. We discuss the interpretation of significances as evidence in \ccite{Fowlie:2019ydo,Fowlie2021Comment}; we focus here only on their correct computation. Although the computation is complicated by the look-elsewhere effect (LEE) and the fact that the signal strength parameters must be positive, the issue with \ccite{Crivellin:2021ubm} is simple: they computed the significances as though they fitted a single signal when in fact they fitted six independent signal strengths. 

\section{Local significance}

To compute significances, it is common to resort to asymptotic formulae (see e.g., \ccite{Cowan:2010js}) based on Wilks' theorem~\cite{Wilks:1938dza}. Roughly speaking, this theorem states that test-statistics of the form \cref{eq:lambda} are asymptotically $\chi^2_k$ distributed, where $k$ is the number of parameters describing the signal. The resonance search, however, violates a required assumption (see e.g., \ccite{Algeri:2020pql}), as the mass parameter isn't identifiable under the background model, i.e., when $\boldsymbol{\mu} = 0$, the model doesn't depend on $m$. 

By simply fixing the mass parameter to its best-fit value $\hat m$, however, we may compute a local \pvalue. In this case, the test-statistic splits into independent contributions for each channel,
\begin{equation}\label{eq:lambda_sum}
    \lambda = \sum_{i=1}^6
       -2 \log{\frac{
        \max_{\mu_i,\theta_i} \like_i(\hat m, \mu_i, \theta_i)
        }{
        \max_{\theta_i} \like_i(\mu_i  = 0, \theta_i)
        }}.
\end{equation}
There is a further complication, though, as the signal strength parameters must be positive, such that the background model lies at the boundary rather than in the interior of the background plus signal model. This is addressed by Chernoff's modification to Wilks' theorem~\cite{10.1214/aoms/1177728725} and further explored in \ccite{10.2307/2289471}. For each channel, if the best-fit signal at a fixed mass fluctuates downwards, that channel's contribution to the test-statistic in \cref{eq:lambda_sum} vanishes. If it fluctuates upwards, that channel's contribution to the test-statistic is described by a chi-squared distribution with one degree of freedom as usual. This is a mixture distribution called a half chi-square distribution, $\halfchisq$. The probability density function is (see case 5 in \ccite{10.2307/2289471} or discussion around eq.~52 in \ccite{Cowan:2010js})  
\begin{equation}\label{eq:half}
    p(\lambda) = \frac{1}{2} \delta(\lambda) + \frac{1}{2} p_{\chi^2_1}(\lambda).
\end{equation}
where $p_{\chi^2_1}(\lambda)$ is the probability density function for a $\chi^2$ distribution with one degree of freedom. The terms in this mixture correspond to downwards and upwards fluctuations and the coefficients are half because downwards and upwards fluctuations are equally likely.

The test-statistic for $n$ channels as in \cref{eq:lambda_sum} for $n = 6$ is thus the sum of $n$ half-chi-squared variates. We denote this as a $\halfchisq_n$ distribution. The density is (see case 9 with $s = 0$ in \ccite{10.2307/2289471}) 
\begin{equation}\label{eq:n}
    p(\lambda) = \left(\frac12\right)^n \sum_{i=0}^n {n \choose i} p_{\chi^2_i}(\lambda),
\end{equation}
where $p_{\chi^2_0}(\lambda) = \delta(\lambda)$. The coefficients are just the binomial coefficients describing the chances of $i$ upwards fluctuations at the fixed mass in $n$ channels. When $i$ of them fluctuate upwards, the test-statistic is a sum of $i$ chi-squared variates which follows a $\chi^2_i$ distribution. When $n = 1$, this reduces to \cref{eq:half}.

In the $n = 1$ case, the density in \cref{eq:half} leads to the simple result for the local significance (eq.~52 in \ccite{Cowan:2010js}),
\begin{equation}\label{eq:Z_local}
    Z_\text{Local} = \sqrt{\lambda}.
\end{equation}
This does not apply when $n \neq 1$ but was used~\cite{email} in \ccite{Crivellin:2021ubm} to calculate $Z_\text{Local} = 5.1\sigma$.
%
In contrast, using \cref{eq:n}, we obtain $Z_\text{Local} = 4.1\sigma$ when combining $n = 6$ channels as in \ccite{Crivellin:2021ubm}.
We validate this result through Monte Carlo (MC) simulations described in \cref{sec:global}. Thus to avoid overstating the significance and reaching potentially faulty conclusions, we must take into account that we searched six channels. 

\section{Global significance}\label{sec:global}

\begin{figure}[t]
\centering
\includegraphics[clip, trim=0cm 0.7cm 0cm 0.6cm, width=0.99\linewidth]{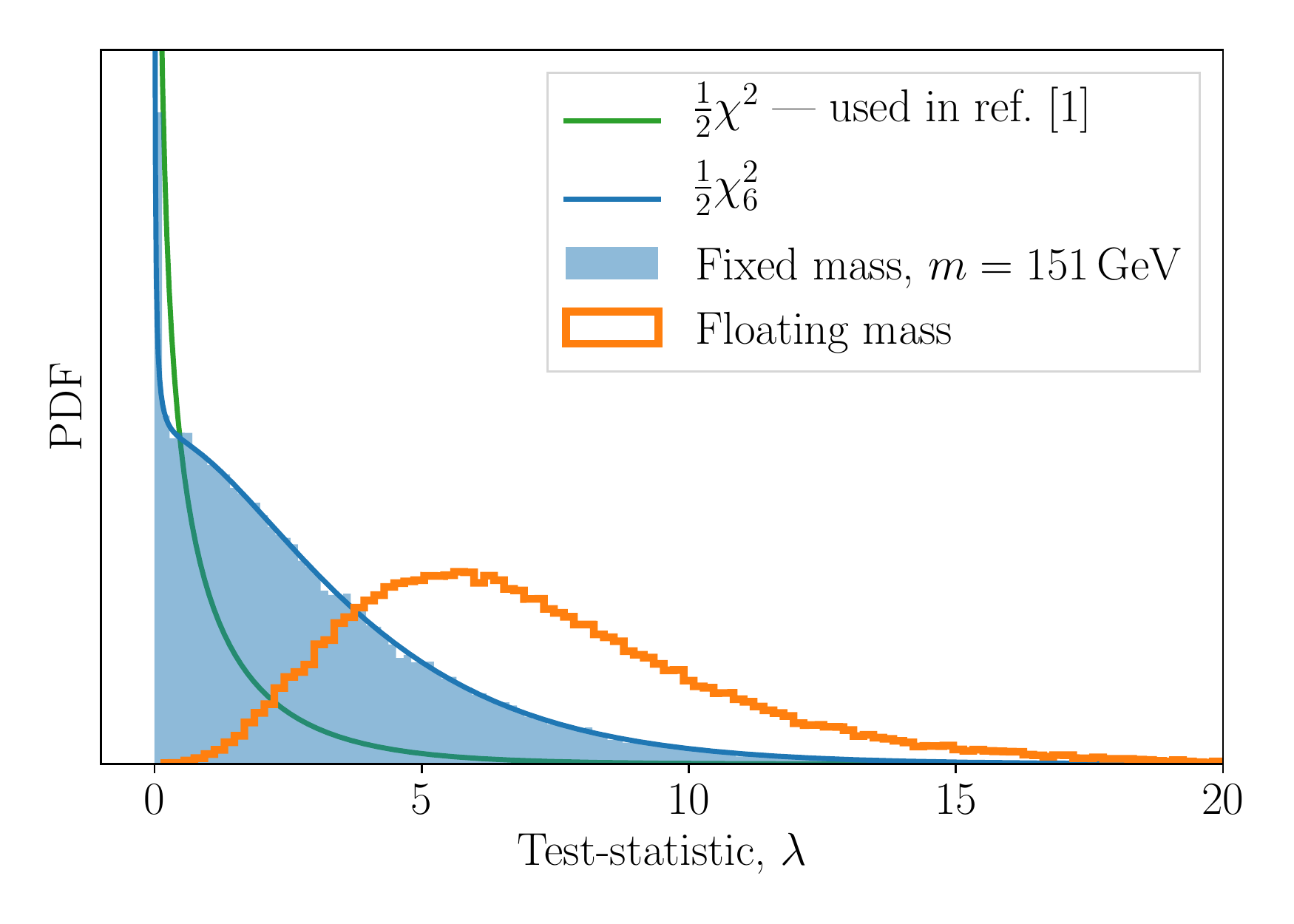}
\caption{Distribution of test-statistics from MC simulations. The $\halfchisq$ and $\halfchisq_6$ distributions are shown by smooth lines for comparison.}\label{fig:dist}
\end{figure}

We previously considered a local significance based on a fixed mass. We should compute a global significance by taking into account every test that was performed and every test that would have been performed were the data different (see problems 1 and 2 in \ccite{Wagenmakers2007} for a pedagogical if partisan discussion). Unfortunately, we don't know the analysis plan of the authors of \ccite{Crivellin:2021ubm} and thus there are imponderable look-elsewhere effects associated with their choices of statistical tests and datasets. Even if the authors of \ccite{Crivellin:2021ubm} began with a particular theory of a new scalar boson in mind, many analysis choices could have been influenced by the observed data. We are not, however, suggesting conscious hacking of any sort and direct readers to the nuanced discussion in \ccite{forking_paths,gelman2014statistical}. 

We don't further discuss these issues and instead compute the global significance following the scope of the LEE considered in \ccite{Crivellin:2021ubm} by taking into account the LEE for the mass range that was searched. In this case, we consider the test-statistic as a random field over the mass range and consider its most extreme fluctuation. \Ccite{Crivellin:2021ubm} accounted for this LEE by multiplying the local \pvalue by a trial factor of 5~\cite{email}. This was an estimate motivated by the $140 \gev$ -- $155\gev$ range and the $1.5\gev$ and $14\gev$ resolutions in the six channels. We instead compute it.

There are rules of thumb for the trial factor based on the range, $\Delta m$, and resolution, $\sigma_m$, e.g.,
\begin{equation}\label{eq:N_thumb}
    N \approx \frac13 \frac{\Delta m}{\sigma_m} Z_\text{Local}.
\end{equation}
This rule yields a trial factor of about 14, but it was found and tested only for the case of $n=1$ channels~\cite{thumb}. Thus rather than using a rule of thumb, we compute it through MC simulations and the Gross-Vitells method~(see e.g.,~\ccite{adler:2007,Gross:2010qma,doi:10.1080/10618600.2019.1677474}). 

The Gross-Vitells method is a sophisticated treatment based on the expected numbers of up-crossings of random fields above specified thresholds.
To apply it, we required the so-called Euler characteristic densities for the $\frac12 \chi^2_6$ random field and an estimate of the expected number of up-crossings at a specified level. From theorem 1 and remark 2 in \ccite{taylor2013detecting}, the Euler characteristic densities for the mixture $\frac12 \chi^2_6$ are just the same mixtures of the known densities for $\chi^2_n$ random fields (see theorem 15.10.1 in \ccite{adler:2007}). To estimate the expected number of up-crossings, we perform simulations using a toy treatment of a resonance problem with five channels with resolution $1.5\gev$ and one channel with resolution $14\gev$ in a window $140 \gev$ -- $155\gev$. In this toy treatment, we fit the mass and signal strengths of a Crystal ball function on top of a fixed background. The resulting test-statistic should obey the same asymptotic distribution as that in the full treatment. We ultimately find a global significance of about $3.5\sigma$, which corresponds to a trial factor of about 12.

To check the asymptotic results for the local and global significances, we perform MC simulations of the test-statistic. We perform $100\,000$ pseudo-experiments with the toy treatment of the problem described above, computing the test-statistics \cref{eq:lambda,eq:lambda_sum} in each case. By simply counting the number of simulations in which the test-statistics exceeded that observed, we again find $4.1\sigma$ local and $3.5\sigma$ global significances, validating our previous results.\footnote{Note though that this does not validate that the asymptotic limit was valid in the real problem, as our toy treatment used large numbers of expected events in all channels.} The distribution of the test-statistic and the test-statistic at a fixed mass of $151\gev$ are shown in \cref{fig:dist}. We see that the simulations in the latter case closely matches a $\halfchisq_6$ distribution, as expected.

\section{Summary}

Taking into account the fact that in \ccite{Crivellin:2021ubm} six signals were fitted rather than one and properly computing the LEE, we find the significances for a new neutral scalar are only $4.1\sigma$ local and $3.5\sigma$ global. Thus while this anomaly may remain intriguing, it isn't as significant as first thought.

\section*{Acknowledgements}

I was supported by an NSFC Research Fund for International Young Scientists grant 11950410509. I thank the authors of \ccite{Crivellin:2021ubm} for time spent clarifying details of their paper.

\newcommand{\JournalTitle}[1]{#1}
\bibliographystyle{references}
\bibliography{references}
\end{document}